\documentclass[english]{revtex4}
\usepackage[LGR,T1]{fontenc}
\usepackage[latin9]{inputenc}
\usepackage{array}
\usepackage{textcomp}
\usepackage{graphicx}

\makeatletter

\DeclareRobustCommand{\greektext}{%
  \fontencoding{LGR}\selectfont\def\encodingdefault{LGR}}
\DeclareRobustCommand{\textgreek}[1]{\leavevmode{\greektext #1}}
\ProvideTextCommand{\~}{LGR}[1]{\char126#1}

\providecommand{\tabularnewline}{\\}

\@ifundefined{textcolor}{}
{%
 \definecolor{BLACK}{gray}{0}
 \definecolor{WHITE}{gray}{1}
 \definecolor{RED}{rgb}{1,0,0}
 \definecolor{GREEN}{rgb}{0,1,0}
 \definecolor{BLUE}{rgb}{0,0,1}
 \definecolor{CYAN}{cmyk}{1,0,0,0}
 \definecolor{MAGENTA}{cmyk}{0,1,0,0}
 \definecolor{YELLOW}{cmyk}{0,0,1,0}
}

\makeatother

\usepackage{babel}
\begin{document}

\title{``Higgs'' Factory at the Greek-Turkish Border\\
{\normalsize{}A Regional Project}}

\author{Serkant Ali \c{C}ET\.{I}N}

\affiliation{\.{I}stanbul Bilgi University, \.{I}stanbul, Turkey}

\author{Evangelos N. GAZIS}

\affiliation{National Technical University, Athens, Greece}

\author{Bora I\c{S}ILDAK}

\affiliation{\"{O}zye\u{g}in University, \.{I}stanbul, Turkey}

\author{Fatih \"{O}mer \.{I}LDAY}

\affiliation{Bilkent University, Ankara, Turkey}

\author{Konstantinos KORDAS}

\author{Chariclia PETRIDOU }

\affiliation{Aristotle University of Thessaloniki, Thessaloniki, Greece}

\author{Yannis K. SEMERTZIDIS}

\affiliation{CAPP, IBS and KAIST, Department of Physics, Daejeon, Republic of Korea}

\author{Saleh SULTANSOY }

\affiliation{TOBB University of Economics \& Technology, Ankara, Turkey}

\affiliation{ANAS, Institute of Physics, Baku, Azerbaijan}

\author{G\"{o}khan \"{U}NEL }

\thanks{contact person, gokhan.unel@cern.ch}

\affiliation{University of California at Irvine, Irvine, USA}

\author{Konstantin ZIOUTAS }

\affiliation{University of Patras, Patras, Greece}
\begin{abstract}
We would like to propose the construction of the photon collider based
\textquotedbl{}Higgs factory\textquotedbl{} in the coming years at
the Greek-Turkish border, starting from its test facility with a high
energy photon beam. This proposal \cite{contribution-espg-2012} was among the contributions to the Open Symposium of the ESPG'12 \cite{about-espg-2012}.
\end{abstract}
\maketitle

\section{Introduction}

After the discovery of the new boson at the LHC \cite{new-boson},
the natural next step is to produce it in a dedicated factory for
detailed studies. There are three possibilities for building such
a factory: 
\begin{description}
\item [{1)}] e$^{+}$e$^{-}$ collider at a center of mass of about 260
GeV to produce ZH final states,
\item [{2)}] a photon-photon collider at the resonance of 126 GeV (corresponding
to E$_{ee}$=160 GeV),
\item [{3)}] a muon collider at the resonance of 126 GeV.
\end{description}
From these three options the last one requires a serious research
and development effort, among other topics on the muon cooling, making
it the least feasible at the present time, considering the current
know-how. Therefore, one really needs to compare the first two options. 

The cross section of the Z associated Higgs production at the e$^{+}$e$^{-}$
collider at $\sqrt{s}$ = 260 GeV with unpolarized beams is 0.2 pb,
about the same as the direct Higgs production at a polarized gamma-gamma
collider. Such a photon collider could be obtained from electron beams
of 80 GeV via Inverse Compton Scattering. The achievable luminosities
are also comparable: $\mathcal{L}=$10$^{33}$ cm$^{-2}$ s$^{-1}$
or better has been previously envisaged \cite{CICHE}. However, the
photon collider option has the advantage of the determination of the
spin and CP properties of the new boson, by adjusting the polarization
of the electron and laser beams. The idea of a photon collider was
first suggested in 1982 \cite{ilk fikir} and later updated for various
electron machines that were planned: for example see \cite{gg@tesla}
and the references therein.

Therefore, we would like to revive the idea of construction of a photon
collider based \textquotedbl{}Higgs factory\textquotedbl{} in the
coming years. We propose South-East Europe, Caucasus and Middle-East
countries as the possible hosts for such a laboratory. Establishment
of such a facility in this area, would help the development of both
accelerator and high energy physics in the region. At the same time
this project would promote peace in this region through international
scientific collaboration. 

Inspiring from the very successful CERN experience, we propose the
Greek-Turkish border in Thrace as the laboratory location.

\section{Electrons and Positrons vs Gammas}

An electron positron collider to study the ZH final states could be
made with either circular or linear machines. The first one has the
obvious advantage of reusing the positrons at the expense of a large
radius of curvature to reduce the energy loss by synchrotron radiation.
For the second case, a fast source has be established to produce positrons
copiously to have the desired luminosity. 

\begin{figure}[h]
\includegraphics[scale=0.42]{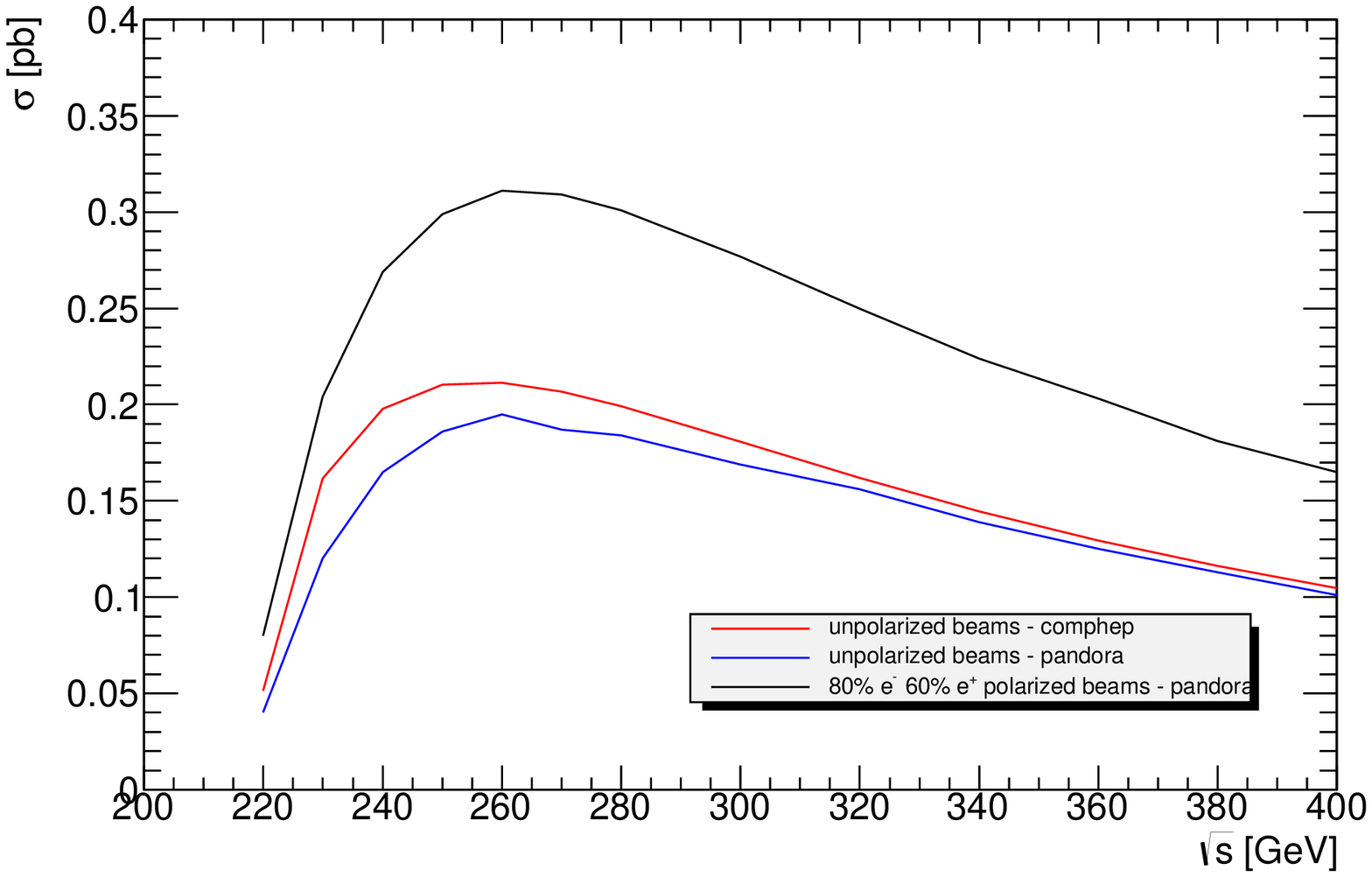}\includegraphics[scale=0.42]{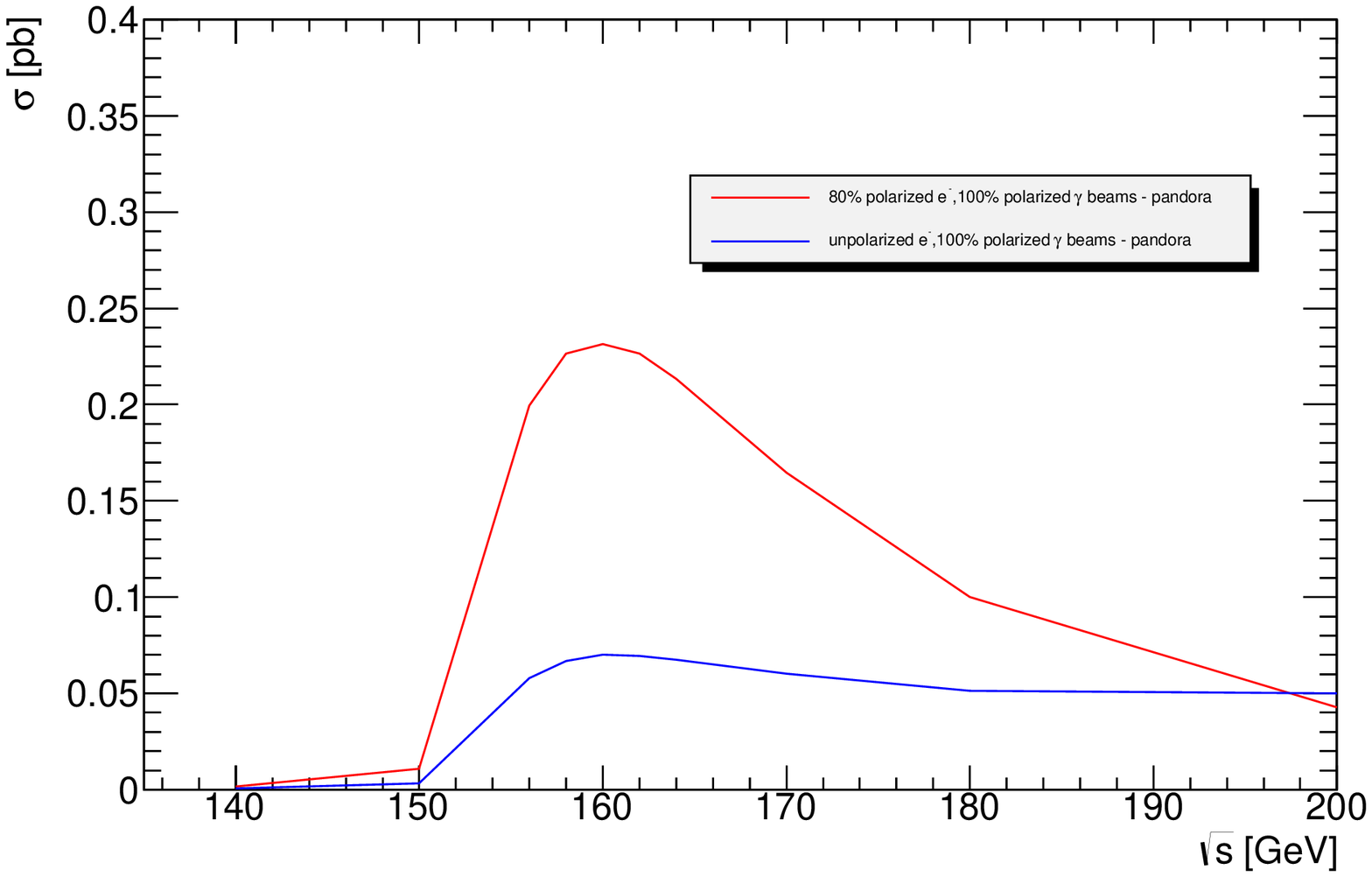}

\caption{Left: Higgs production cross section in an e$^{+}$e$^{-}$ collider
as a function of $\sqrt{s}$ for the ZH final state. Right: Higgs
production cross section in a $\gamma\gamma$ collider as a function
of $\sqrt{s}$ (of the $e^{-}e^{-}$ machine) to produce H bosons
in he final state. Two different numerical tools are used to consider
both polarized and unpolarized beams\label{fig:Higgs-ee}.}

\end{figure}

The cross section for the ZH final state has been calculated by using
two independent tools: CompHEP \cite{comphep} and Pandora \cite{pandora}.
The results are shown in Fig. \ref{fig:Higgs-ee} left side, with
both unpolarized and polarized electron beams. Following the earlier
examples in the field, an optimistic scenario is considered: the electron
(positron) polarization is assumed to be 80\% (60\%). In this case,
the maximum cross section is 0.20 pb for the unpolarized and 0.31
pb for the polarized beams.

A similar plot for proposed $\gamma\gamma$ collider can be found
in Fig. \ref{fig:Higgs-ee} right side. The maximum cross section
that can be achieved is about 0.23 pb with polarized electron beams
and assuming 100\% laser polarization. Similar to previous case, the
electron polarization is taken to be 80\%. One should note that, for
such a machine, one doesn't need to produce positrons. Previous work
has shown that in a $\gamma\gamma$ collider, the achievable luminosities
are better than 1/10 of the luminosity that can be obtained from an
$e^{-}e^{-}$ machine, meaning somewhere between 10 to 100 fb$^{-1}$.
Such a collider would therefore yield, 2300 to 23000 Higgs bosons
per year.

A $\gamma\gamma$ machine would also have the intrinsic capability
to produce $\gamma e^{-}$ collisions with a considerable physics
potential on its own. For example, it provides an opportunity to measure
the triple gauge boson coupling WW$\gamma$, without involving the
ambiguities from quartic or ZWW couplings \cite{WWg}.

\section{Photon Collider Specifics}

\subsection{Accelerator Considerations}

A few words about the photo-injectors \& fiber lasers, including the
photo-cathode shape.

Utilisation of a microtron a the second stage should reduce the overall
length of the machine.

A dedicated electron linac with two arcs bending in opposite directions
or two independent electron linacs facing each other seem to be the
two possible options for obtaining the 80 GeV electron beams. The
relative merits of both approaches have been previously considered
within the context of CLIC1 with the conclusion that the first option
was simpler and cheaper to build, whereas the second one was preferable
from a performance point of view \cite{Schulte04}. The utilization
of CLIC technology would permit an accelerating gradient of about
100 MV/m, thus an electron machine of about 1.5 km would reach the
target energy of 80 GeV. For comparison, the ILC design with superconducting
cavities yields a gradient of about 35 MV/m thus roughly tripling
the length of the accelerator. The CLIC option has the potential of
being cheaper since apart from being shorter, it uses normal conducting
accelerating structures.

CLIC accelerating gradient is feasible today? They didn't progress
much.

High energy photon beam will be obtained via Compton backscattering
of a laser beam off the electron beam at Conversion Point (CP) which
is prior to interaction point of $\gamma\gamma$ collisions. The distance
between the CP and the interaction region could be as small as 5 cm. 

\subsection{Laser Technologies}

The technological basis of generation of highly intense ultrafast
pulses is rapidly developing on multiple fronts and a future laser-based
source for gamma particles is likely to be well-served by these advances.
The requirements for high flux translate into large average powers,
as well as high peak powers and short pulses to maximize the interaction
cross section for (inverse) Compton scattering, which appears to be
most feasible approach to laser-based gamma ray generation.

The wavelength needed to obtain a $\gamma$ beam of 64 GeV can be
easily found as $\lambda=0.38$ $\mu$m according to the well known
formula: $E_{\gamma}^{max}=E_{e}\frac{x}{x+1}\,,\;\;x\approx19\times E_{e}[TeV]/\lambda[\mu m]$
as given in \cite{gg@tesla}. As such a wavelength is hard to obtain
for extremely high average power and pulse energy combinations required,
a tradeoff can be envisaged to slightly increase the electron beam
energy to $E_{e}=85$ GeV, which would require a longer laser wavelength
to match the change in the $x$ factor in the maximum photon energy
calculation. With $\lambda=0.53$ $\mu$m, $x$ becomes 3.2 yielding
the required $E_{\gamma}^{max}=64$ GeV. The main parameters of the
proposed photon collider are summarized in Table \ref{tab:machine-params}.

There are three potential technology lines that may be pursued to
achieve the desired laser parameters: (1) traditional solid state
laser technology, (2) thin-disk lasers, and (3) fiber lasers. In all
three cases, conversion to 0.5 $\mu$m region is accomplished via
second harmonic generation (SHG).
\begin{enumerate}
\item Solid state lasers: Well-established ultrafast solid state lasers
(most notably Ti:sapphire, Nd:glass, etc) have generated even J-level
energies with picosecond and femtosecond pulse durations being routine.
Nevertheless, it will be challenging to scale up these lasers to kW-level
average powers. One notable exception is cryogenically cooled Yb:YAG,
with which 2.3 kW CW operation has been demonstrated \cite{FOI1}.
In the pulsed regime, 40 mJ pulses at 1 kHz and 15 ps has been generated
\cite{FOI2}. However, the narrow bandwidth of the gain medium limits
pulse durations to above few ps or longer.
\item Thin disk lasers: There has been significant advances in this area,
with already >1 kW-level average powers and high energies being demonstrated
\cite{FOI3}. Thin-disk lasers hold good potential for scaling up
power levels even further. However, these systems are quite complex
at high powers and pulse scaling has been relatively little studied
for short pulses.
\item Fiber lasers: A non-conventional approach would be to capitalize on
the developments in fiber lasers, which have excellent performance
at high average powers, with as much as 10 kW demonstrated from a
nearly diffraction-limited CW source \cite{FOI4}. Even with sub-picosecond
pulses, 830 W has been generated from rod-type fiber lasers \cite{FOI5}.
While fiber lasers have practical advantages, such as robustness,
lower cost, etc, their outstanding drawback is limited potential for
scaling up the pulse energy due to peak limitations induced by nonlinear
effects. However, this limitation can be circumvented through the
use coherent beam combining. Given the relative simplicity and very
little added cost of developed multiple fiber amplifier branches of
total power equal to that of a single amplifier, and recent advances
in coherent beam combining, this is an attractive prospect \cite{FOI6}.
One could imagine combining up to 100 such sources into a single,
massive powerful beam. An additional prospect is the equally rapid
development of burst-mode fiber lasers, which generate a group of
closely spaced high-energy pulses \cite{FOI7,FOI8}.
\end{enumerate}
Here, we would like to express particular interest in meeting the
challenging laser technology requirements through the development
of fiber laser technology given their potential to surpass the other
alternatives in a $\sim$5-year time frame, although it is evident
that competing technologies should also be carefully considered and
evaluated. With the fiber approach, pulse energy can potentially be
scaled up to 1-mJ for all-fiber-integrated amplifiers and few-mJ through
the use of specialty rod-type amplifiers, assuming, in both cases,
use of phase compensation techniques. 

Massively beam combining burst-mode ultrahigh-power fiber amplifiers:
Considering rapid developments of coherent beam combining of 100-1000
W level CW fiber lasers, it is reasonable to expect that this technology
will mature in the 4-5 year time scale, allowing combining potentially
more than 100 beams. Although this brings up increased complexity,
the potential for all-fiber-integrated amplifiers greatly simplify
the laser design. Therefore, it seems reasonable to propose development
of a unique laser system, comprising of 100 amplifier channels, each
providing, e.g., 1 mJ/pulse, with 100 mJ/burst at 1 kHz and 100 W
average power at 1.06 \textgreek{m}m. Thus, the combined system, after
SHG, could offer 50 mJ/pulse with 5 J/burst at 1 kHz and 5 kW average
power at 530 nm. Pulse durations in the range of 1 ps could be targeted.

\begin{table}[h]
\caption{The parameters of the proposed $\gamma\gamma$ collider }
\label{tab:machine-params}
\centering{}%
\begin{tabular}{c|c}
Parameter & Value\tabularnewline
\hline 
E$e^{-}$ (GeV) & 85 (80)\tabularnewline
\hline 
E$\gamma$ (GeV) & 64\tabularnewline
\hline 
$\mathcal{L}$$_{\gamma\gamma}$ (10$^{33}$ cm$^{-2}$ s$^{-1}$) & 1 .. 10\tabularnewline
\hline 
Laser wavelength ($\mu m$) & 0.53 (0.38)\tabularnewline
\hline 
repetition frequency (Hz) & \multicolumn{1}{c}{same as electron beam}\tabularnewline
\end{tabular}
\end{table}

\subsection{Other Physics Potential}

Apart from the already mentioned studies on the ``Higgs'' properties
which would give the discriminating power between various models (such
as MSSM Higgs vs SM Higgs) and the W-W-$\gamma$ coupling studies,
a 160-GeV photon collider would have possibility for other interesting
physics cases as well. For example the photon structure function could
be investigated with the $\gamma e$ collisions, especially the gluon
distribution of the photon. The possibility to change the polarization
of the beams would also allow to determine the spin dependent structure
function of the photon \cite{gg@tesla}. Another interesting case
is the study of the W boson in detail. It could be single produced
in the $\gamma e$ collisions and pair produced in the $\gamma\gamma$
collisions, where the latter at $\sqrt{s}=160$ GeV would yield very
few events (about 24 per year in the best case) but would also permit
the investigation of the WW$\gamma$$\gamma$ quartic vertex. However
increasing the electron beam energy by 10 GeV would increase the yearly
event yield 10 fold. 

\subsection{Location Considerations}

\begin{figure}[h]
\includegraphics[scale=0.3]{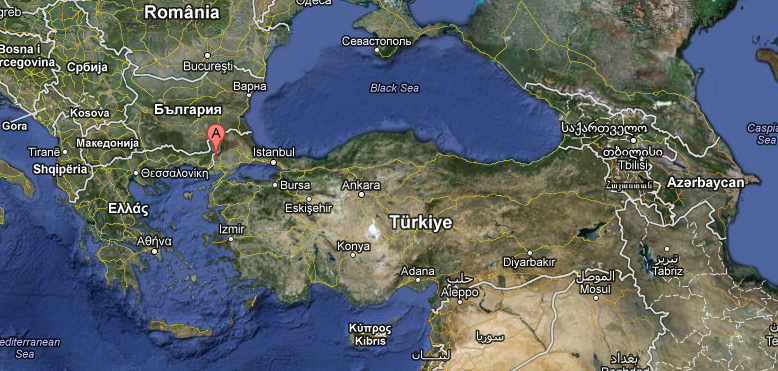}\  \includegraphics[scale=0.2]{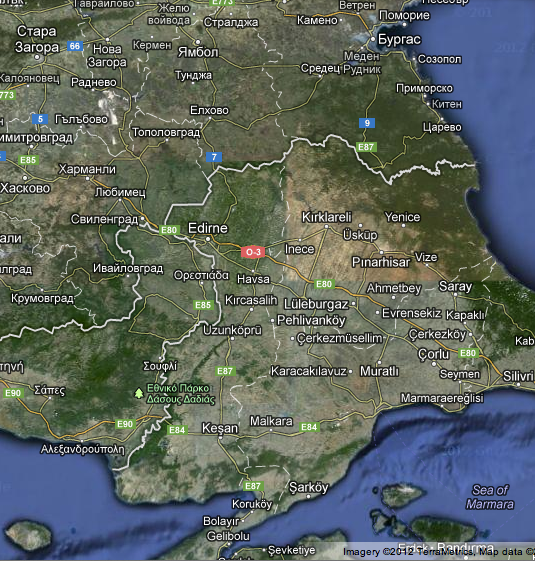}

\caption{Geographical location of the suggested laboratory. The Greek-Turkish
border is marked on the left side and zoomed in on the right side.
\label{fig:Geo}}
\end{figure}

We propose the establishment of a regional laboratory in Thrace to
promote the accelerator and high energy physics potential of South-East
Europe, Caucasus and Middle-East. The Fig. \ref{fig:Geo} contains
a map of the region with a marker showing the possible laboratory
site. If the two linac option of the machine is adopted, Greece and
Turkey could host one linac each and the detector(s) could be right
at the border. If a single linac with two arcs is selected the linac
could be along the borderline, with facilities on each side of the
border.

Thrace region has a low cost of real estate, manpower and living.
Therefore the overall project would require lower investments compared
to a similar installation at CERN or at any other Western European
location. Additionally the local industries would benefit from this
project and the region would greatly develop, leading to an increased
gross regional productivity.

\section{Conclusions}

We believe the world would benefit from a photon collider for a multitude
of reasons: foremost, a detailed study of the boson that has recently
been discovered at LHC is needed. The branching fractions, the CP
properties would be investigated in details. Additionally the $\gamma e$
collisions would be available, opening the potential for other physics
studies such as triple gauge boson couplings.\emph{ }

We believe the South-East Europe, Caucasus and Middle-East region
could benefit from an international laboratory for a multitude of
reasons: foremost, the establishment of such a facility in this area
would promote peace in this region through international scientific
collaboration. Inspiring from the very successful CERN experience,
a double or triple country border could be the the ideal laboratory
location. Additionally it would also help the development of both
accelerator and high energy physics as well as the laser technology
in this part of the world.

In conclusion, we propose the construction of a photon collider based
\textquotedbl{}Higgs factory\textquotedbl{} as an international collaboration
in Thrace, where Greece and Turkey (and even Bulgaria) border each
other. The very first milestone towards such an effort is a good understanding
of the Conversion Point. The establishment of a regional laboratory
(with the prospect of enlargement), with the precise initial target
of obtaining a high energy photon beam is the very first step towards
the final goal.

\end{document}